\documentclass[11pt, oneside]{article} 

\usepackage{hyperref}
\usepackage{geometry}                		
\usepackage{graphicx}														
\usepackage{amssymb}
\usepackage{amsmath, amssymb, graphics, setspace}
\newcommand{\mathsym}[1]{{}}
\newcommand{\unicode}[1]{{}}

\usepackage{amsmath}
\usepackage{wrapfig}    
\usepackage{graphicx}
\usepackage{amsthm}
\usepackage{bbm}
\usepackage[dvipsnames]{xcolor}
\usepackage{bm} 
\usepackage{appendix}

\newtheorem{prop}{Proposition}

\usepackage[utf8]{inputenc}
\usepackage[dvipsnames]{xcolor}
\usepackage{fancyhdr}
\usepackage{mathrsfs}
\newtheorem{definition}{Definition}[section]

\title{A Multivariate Hawkes Process Model for Stablecoin-Cryptocurrency Depegging Event Dynamics}
\author{Connor Oxenhorn}
\date{April 18, 2022}
\begin{document}
\maketitle
\thispagestyle{fancy}

\begin{abstract}
Stablecoins—digital assets pegged to a specific currency or commodity value—are heavily involved in transactions of major cryptocurrencies \cite{treasury2021stablecoins}. The effects of deviations from their desired fixed values (depeggings) on the cryptocurrencies for which they are frequently used in transactions are therefore of interest to study. We propose a model for this phenomenon using a multivariate mutually-exciting Hawkes process, and present a numerical example applying this model to Tether (USDT) and Bitcoin (BTC).
\end{abstract}
\section{Introduction}
\subsection{Stablecoins and Cryptocurrency}
The rise of digital markets—whether with respect to the digitalization of equity markets in the early 2010's or the increasing adoption of digital assets in the early 2020's—always brings with it new dynamics that are important for market participants to identify and understand. 
\par
Stablecoins are digital assets, predominantly supported by public blockchain networks, which are designed with the purpose of maintaining a stable value relative to a reference asset. These reference assets are usually national currencies or commodities. Stablecoins are primarily used to facilitate trading and lending of other digital assets, as well as to move funds easily between digital asset platforms. In the U.S., most stablecoins are considered ``convertible virtual currency" meaning that they serve as substitutes for a specific reference asset \cite{treasury2021stablecoins}. In the case where a stablecoin is not convertible for an underlying reference asset, it achieves its desired fixed value using a seigniorage algorithm, which buys and sells in the open market to maintain this value \cite{zhao2021understand}. In either case, due to external market forces, it is easy to see that a stablecoin's price may deviate considerably from its desired value—an event referred to as a depegging.
\par
The use of stablecoins in the buying and selling of major cryptocurrencies has increased dramatically in recent years. For example, some blockchain research firms estimate that stablecoins, namely Tether (USDT), are involved in over 70 percent of Bitcoin (BTC) transactions (as per Fall of 2021) \cite{kaikobitcointether}. Thus, during depegging events, we expect there to be ripple effects in the microstructure of cryptocurrency markets.
\subsection{Hawkes Process}
We assume that stablecoin depeggings can be characterized as excitatory events. More rigorously, this means that an occurrence of a depegging increases the probability of another depegging in the near-term. Additionally, since we are interested in modeling the interactions between stablecoin depeggings and disruption events in the price of a given cryptocurrency, we assume that the occurrence of a depegging also increases the probability of a disruption event in the price of a given cryptocurrency in the near-term. Based on these assumptions, any model proposed for this phenomenon must be capable of capturing the above, as well as any potential effects on the stablecoin depeggings due to disruption events in the cryptocurrency price (although we anticipate these interactions to be fairly one-sided).
\par
Since we are interested in studying this interaction through the scope of `events', i.e. depeggings and disruption events in cryptocurrency price (granular price jumps), we look to a class of stochastic point processes called counting processes. Given that we would like to capture excitatory dynamics between these two types of events, we employ an excitatory counting process that can be easily generalized to the multivariate case, which is of particular interest to us, called a Hawkes process. Originally used for applications in seismology, Hawkes processes have found numerous applications in high-frequency trading, social network theory, and many other fields \cite{lemonnier2017multivariate}. These processes incorporate parameters that can capture excitation dynamics pertaining to both magnitude and duration of self- and cross-excitation between different dimensions (types) of events. 
\par
For this reason, we choose to construct our model using a multivariate Hawkes process, which will be defined below.

\begin{definition}[Stochastic Process]
Let $(\Omega, \mathcal{H}, \mathbb{P})$ be a probability space, with sample space $\Omega$,  $\mathcal{H}$ a $\sigma$-algebra on $\Omega$, and probability measure $\mathbb{P}: \mathcal{H}\to [0,1]$. Additionally, let $X:\Omega\to\mathbb{R}$ be a real-valued random variable. Then we call the collection of random variables $\{X_i : i\in\mathbb{N}\}$ a stochastic process.
\end{definition}
\par
For a random variable $X$ defined on the space $(\Omega, \mathcal{H}, \mathbb{P})$, we refer to a collection $\{\mathcal{H}_t\subseteq\mathcal{H}:t\geq 0\}$ with the property $\mathcal{H}_t\subseteq\mathcal{H}_u$ iff $t\leq u$ as a \textit{filtration} of $X$. Furthermore, any specific $\mathcal{H}_t$ is referred to as the \textit{filtration of X up to time t}.

\begin{definition}[Counting Process]
Let $\mathcal{N} = \{N(t)\in\mathbb{N} : t \geq 0\}$ be an almost surely finite stochastic process where $N(0)=0$ and its jump trajectories are right-continuous step functions with increments of one unit. Then, we call $\mathcal{N} $ a counting process.
\end{definition}
\par
We say $\{t_i>0 : i\in\mathbb{N}\}$ is the set of \textit{arrival times}, where each $t_i$ is the time of the $i^{th}$ jump. With this notation, we provide an alternate formulation for random variables in a counting process: $$N(t) = \sum_{i\in\mathbb{N}} \mathbbm{1}_{t_i\leq t}.$$ 

\begin{definition}[Conditional Intensity Function]
Take $\mathcal{N}$ to be a counting process and $\mathcal{H}_t$ to be the filtration of $\mathcal{N}$ up to time $t$. We define the conditional intensity function of $\mathcal{N}$ as $$\lambda(t) = \lim_{h\to0^+} \frac{\mathbb{E}[N(t+h)-N(t)|\mathcal{H}_t]}{h}.$$
\end{definition}
For the class of counting processes we are interested in for this paper, our conditional intensity functions take the form $$\lambda(t) = \mu + \int_0^t f(t-x)dN(x)$$ where $\mu > 0$ is a \textit{background intensity} constant and $f:(0,\infty)\to(0,\infty)$ is an \textit{excitation function}, also referred to as a \textit{kernel function}.

\begin{definition}[Multivariate Hawkes Process]
Consider the collection of $m$ counting processes $\mathcal{S} = \{\mathcal{N}_1(t), \mathcal{N}_2(t), \dots, \mathcal{N}_m(t)\}$ with a  set of arrival times for each counting process $\{t_i^{(j)} : i\in\mathbb{N},j=1,2,\dots,m\}$ and a set of filtrations $\{\mathcal{H}_t^{(j)} : t\geq0, j=1,2,\dots,m\}$. Suppose that for some $h>0,t\geq 0$ each counting process $\mathcal{N}_j(t) \in\mathcal{S}$ has the property  $$\mathbb{P}(\mathcal{N}_j(t+h)-\mathcal{N}_j(t)=n|\mathcal{H}_t^{(j)}) = 
\begin{cases}
\lambda_j(t)h+o(h), & n=1\\
o(h), & n > 1 \\
1-\lambda_j(t)h+o(h), & n=0
\end{cases}$$ with conditional intensities $$\lambda_j(t) = \mu_j + \sum_{k=1}^m\int_0^t f_{jk}(t-x)dN_k(x)$$ for some $\mu_j \in \{\mu_j >0 : j=1,2,\dots,m\}, f_{jk}:(0,\infty)\to(0,\infty)\in \{f_{jk}: j,k =1,2,\dots,m\}$. Then, $\mathcal{S}$ is a multivariate Hawkes process (also referred to as an $m$-variate Hawkes process). 
\end{definition}
\section{Proposed Model}
\paragraph{Related Work}
There is a notable breadth of work revolving around applications of Hawkes processes to microstructures of asset markets, e.g. modeling order book dynamics \cite{chen2017modelling}. There has also been work applying Hawkes processes to equity and cryptocurrency time series to capture self- and cross-excitation dynamics \cite{bacry2015hawkes}, even generalizing these models to frameworks designed to predict market crashes \cite{gresnigt2015interpreting}. However, to the best of our knowledge, there is little-to-no work utilizing Hawkes processes to model dynamics of the market microstructure between stablecoins and cryptocurrencies.
\subsection{Model Construction}
We begin by specifying magnitude thresholds for the stablecoin depeggings and cryptocurrency price disruption events that we want to track. Note that it may be prudent to use different thresholds for each type of event. We then track these events, recorded as the sets of arrival times $\{t_i^{(s)} : i\in\mathbb{N}\}$, $\{t_i^{(c)} : i\in\mathbb{N}\}$ using a collection of counting processes $\mathcal{S} = \{\mathcal{N}_s(t), \mathcal{N}_c(t)\}$, for depeggings and price jump events in the stablecoin and cryptocurrency respectively.
\par
 A brief note on notation: we use subscripts $s,c$ to denote whether a variable pertains to the time series of either the stablecoin or the cryptocurrency. In cases where variables represent cross-behavior between the two time series, we use the subscript $sc$, $cs$ to denote an effect on the stablecoin time series from the cryptocurrency time series, and vice versa.
\par
To allow this model to capture excitatory behavior between events in the two time series, we choose a common self-exciting kernel function, and apply it to the multivariate case. Let $\alpha,\beta,t>0$ and $x\geq0$, we define the self-exciting kernel function as $$f(t-x) = \alpha e^{-\beta(t-x)}.$$ Here, $\alpha$ captures the sensitivity of the conditional intensity function to new events in terms of magnitude, and $\beta$ captures information about how the conditional intensity function decays after an event. Applying this to the multivariate case of our model, for $\mu_j>0, j\in\{s,c\}$, we get $$\lambda_j(t) = \mu_j + \sum_{k\in\{s,c\}}\int_0^t \alpha_{jk} e^{-\beta_{jk}(t-x)} dN_k(x) = \mu_j + \sum_{k\in\{s,c\}}\sum_{t^{(k)}_i < t} \alpha_{jk} e^{-\beta_{jk}(t-t^{(k)}_i)}.$$
\par
Note that this construction can be generalized to any number of stablecoins and cryptocurrencies, not just two.
\subsection{Log-Likelihood}
The parameters for Hawkes process models are found through maximum likelihood estimation. We present the log-likelihood function for the general $m$-variate case \cite{chen2016likelihood}, as the 2-variate case of our model can be inferred from it.
\begin{prop}
Let $\mathcal{S}$ be an $m$-variate Hawkes process with arrival times $0\leq t_1^{(j)}\leq t_2^{(j)},\dots\leq T$  $\forall j=1,2,\dots,m$. Additionally, let $\theta = \big(\{\alpha_{jk} : j,k=1,2,\dots,m\}, \{\beta_{jk} : j,k=1,2,\dots,m\}, \{\mu_j : j=1,2,\dots,m\}\big)$ be a vector of model parameters such that $\alpha_{jk},\beta_{jk},\mu_j>0$. Then, we define the log-likelihood function of $\mathcal{S}$ to be $$logL_{\mathcal{S}}(\theta) = \sum_{j=1}^m \large(-\mu_jT - \sum_{k=1}^m \frac{\alpha_{jk}}{\beta_{jk}} \sum_{\{t^{(k)}_i:i\in\mathbb{N}\}}(1-e^{-\beta_{jk}(T-t^{(k)}_i)}) + \sum_{\{i:t^{(j)}_i<T\}}\log\large[\mu_j+\sum_{l=1}^m\alpha_{jl}R_{jl}(i)\large] \large),$$ where $$R_{jl}(i) = e^{-\beta_{jl}(t^{(j)}_i-t^{(j)}_{i-1})}R_{jl}(i-1) + \sum_{\{n : t^{(j)}_{i-1}\leq t^{(l)}_n<t^{(j)}_i\}}e^{-\beta_{jl}(t^{(j)}_i-t^{(l)}_n)}$$ with the initial condition that $R_{jl}(0)=0$.
\end{prop}
\section{Numerical Example}
To provide further insight into the model described above, as well as to illustrate some of the benefits and drawbacks of its use in practice, we apply it to an examination of the depegging dynamics between the stablecoin Tether (USDT), and the cryptocurrency Bitcoin (BTC).
\subsection{Discussion of Data}
Due to the fact that depeggings of stablecoins often occur rapidly and are swiftly corrected (either by a seigniorage algorithm or by buyers and sellers in the open market), this model is most appropriately applied to highly granular data. Thus, tick data summarized into one-minute increments is used for both USDT and BTC. Given that we are interested in price disruption events that occur within each minute, we use the $\text{high}-\text{low}$ for each minute interval, standardized into a percentage change.
\par
The granularity of this data introduces computational challenges, and so for this analysis we focus on a specific period of interest—January 19, 2018—during which 96, $90^{th}$-percentile (calculated over the previous 5-year period) depeggings of USDT occurred. The arrival times for these price jump events are formatted into hour units.
\begin{center}
\includegraphics[scale=0.4]{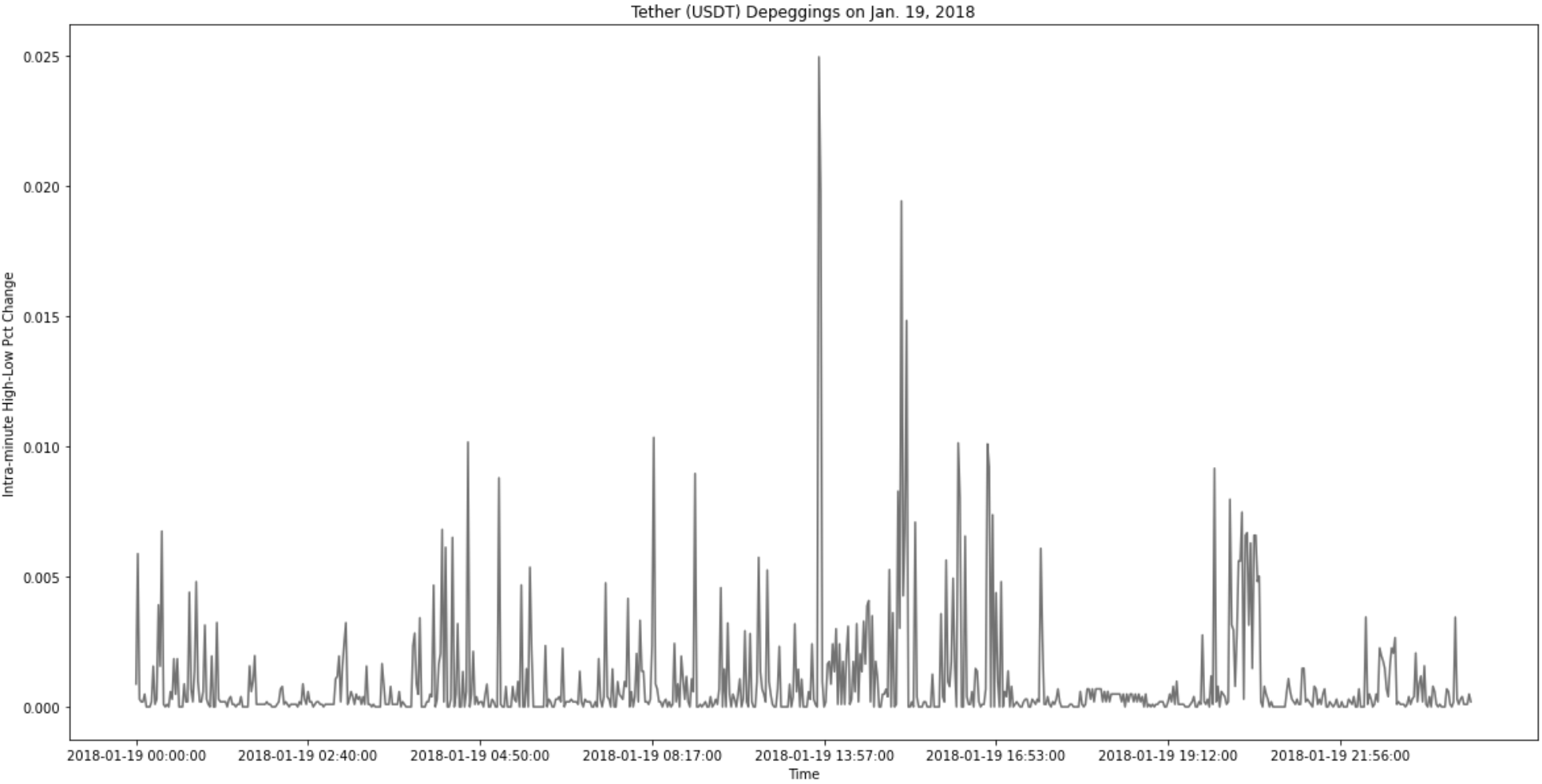}
\includegraphics[scale=0.4]{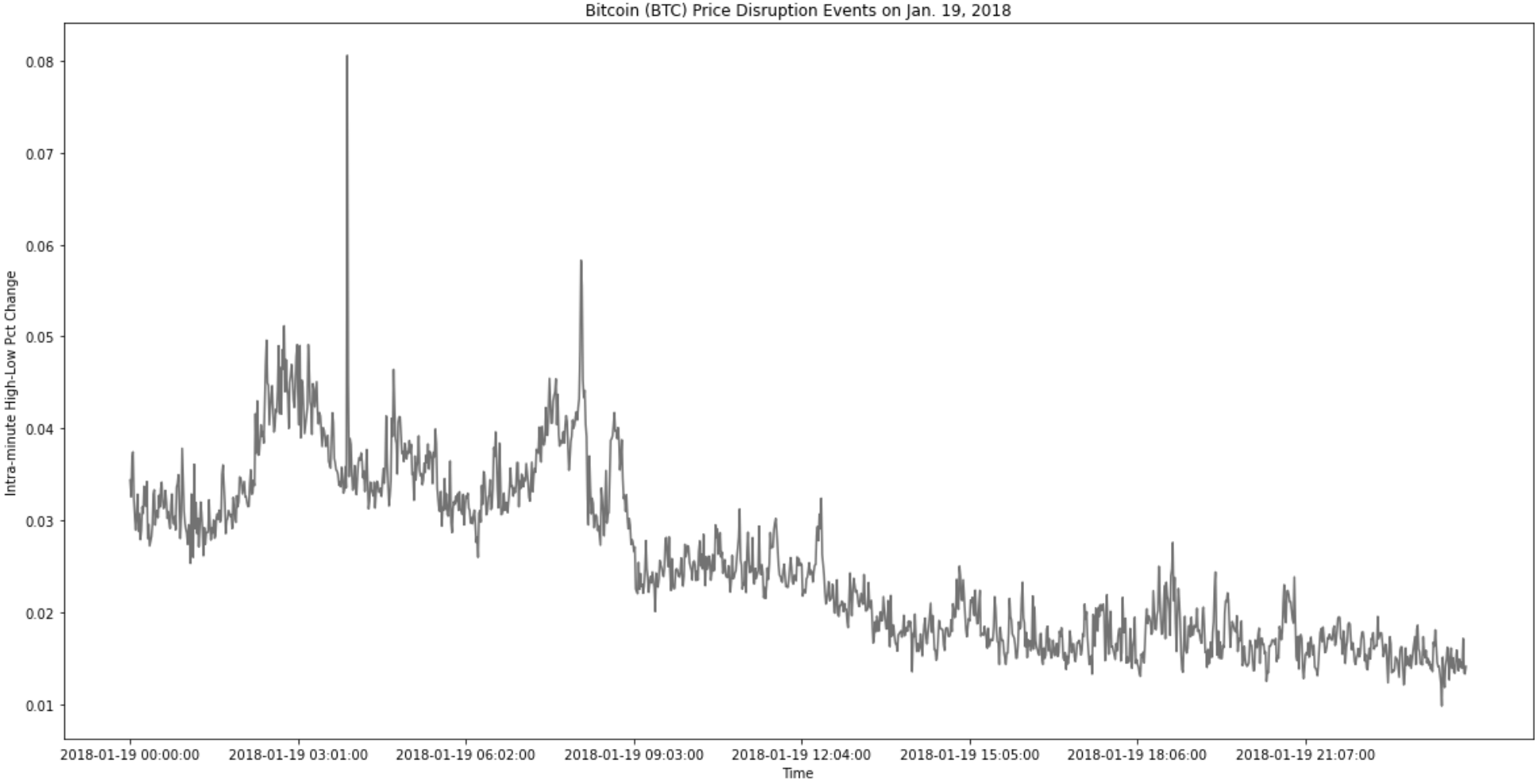}
\end{center}
Another concern with respect to data selection and formatting has to do with the magnitude of price jump events in the BTC timeseries that correspond to (or are caused by) depeggings in USDT. Stablecoin depeggings are events that likely affect the micro structure of cryptocurrency markets. Since the market value of BTC is subject to a plethora of variables—many of which will likely have a significantly larger effect on BTC value than depeggings of USDT will have—we will fit our model to a range of percentiles of BTC price jump events throughout our period of examination, while only using $90^{th}$ percentile USDT depegging events (calculated from the data on the date of examination), and analyze the values of the parameters of our model fitted to each. 
\par
Given that January 19, 2018 was a fairly quiet day in terms of global and national news, we do not anticipate any notable bias in our sample toward outside events that may significantly impact our results.
\subsection{Results}
The estimates for the parameters of our model, for each percentile range of price jumps in the BTC time series, are largely as expected (see Appendix). We see that estimates for cross-excitation parameters based on our data sample are largely asymmetric. For lower percentiles of BTC price jumps, $\hat{\alpha}_{sc} < \hat{\alpha}_{cs}$ and $\hat{\beta}_{sc} > \hat{\beta}_{cs}$. This would suggest that in our data sample, depeggings in USDT have a larger excitatory effect on BTC price jump events than BTC price jump events have on USDT depeggings as well as that a depegging in USDT has a longer duration excitation effect (the conditional intensity decays slower) on BTC price jump events than BTC price jump events have on USDT depeggings. 
\par
The above intuitive behaviors seem to waver at higher percentile ranges of our data. While further analysis is encouraged, this could point to behavior where BTC price jump events that may be caused by USDT depegging events may more commonly be of lower magnitudes; in the case of our data sample: the $10^{th}$-$40^{th}$-percentile range.
\subsection{Optimization}
In order to approximate the parameter values for the model, we must maximize its Likelihood function, or equivalently its log-Likelihood function. We employed the Nelder-Mead simplex algorithm \cite{olsson1975nelder} in doing so. While this method may converge more slowly than a gradient method, as well as may converge to non-stationary points in some instances, we felt this to be most appropriate for two reasons. The first is that calculating the log-Likelihood function—let alone any of its derivatives—in conjunction with the sheer scale of our data, requires significant time and space complexity; so a procedure that simply recursively evaluates the function helps to lessen the computational burden. The second is that during optimization, domain constraints must be placed to ensure the range of the log-Likelihood function remains real. While there are techniques to deal with this sort of issue in gradient optimization, it complicates the optimization procedure.
\subsection{Remarks}
Maximum likelihood estimation for Hawkes processes poses a particularly challenging problem computationally, as even a 2-variate mutually-exciting Hawkes process model relies on 10 individual parameters. In the most convenient case, where our model is fitted to two starkly similar time series, the generous assumptions of approximate symmetry in cross excitation (i.e. $\alpha_{jk}=\alpha_{kj}, \beta_{jk}=\beta_{kj}$) and approximate equivalence in background intensities could be justified. Even in this case, however, the model still relies on 7 individual parameters; and given the granularity of time series data that this model should be fit to, this requires significant computing power. Additionally, bootstrapping methods for initial exact parameter estimation that, for many other statistical models, are effective in decreasing computation time, are inappropriate for Hawkes processes as they rely on the entire history of the process \cite{lemonnier2017multivariate}. In light of this, there are a few areas of interest for further exploration in terms of improved application of the proposed model.
\par
The first has to do with the accuracy of parameter estimation. While the quantity of data involved in fitting these models is admittedly very large, to improve finite sample performance, a bootstrap inference method may be employed. Bootstrap inference procedures for Hawkes processes can consist of asymptotic confidence intervals, or in cases where arrival times occur on the interval $[0,T]$ where $T$ is not sufficiently large, polyhedral confidence sets may be used \cite{wang2020uncertainty}. There also include resampling techniques such as Fixed- or Recursive-Intensity Bootstrap procedures \cite{cavaliere2022bootstrap}.
\par
Another area of interest, which is more so relevant if one were to generalize this model to be $m$-variate for $m$ sufficiently large, includes methods for low-rank approximations of the kernel matrix $\{ f_{jk}:j,k=1,2,\dots,m\}$ \cite{lemonnier2017multivariate}. This can significantly decrease runtime of maximum likelihood estimation procedures, as the dimensionality of the kernel matrix is collapsed with minimal loss of information. In the above case, we have low dimensionality ($m=2$) and the issue lies more in the amount of data and arrival times that need to be considered in parameter estimation. In this case, there are various expectation-maximization (EM) schemes \cite{lewis2011nonparametric} that may be appropriate and worth further exploration.  
\section*{Acknowledgement}
I would especially like to thank my advisor Liyang Zhang for his constant feedback, support, and encouragement toward creativity during this research. I would also like to thank Renato Mirollo, Paul Garvey, Elisenda Grigsby, and the Boston College Mathematics Department for enabling me to pursue this research and providing feedback at critical points of the process. Finally, I would like to thank my good friend Patrick Bjornstad for his suggestions regarding research direction and advice on convergence, runtime, and data formatting.
\bibliography{citations}{}
\bibliographystyle{ieeetr}
\section*{Appendix}
\appendix
\section{\\Parameter Estimates}
\begin{center}
\begin{tabular}{|| c | c c c c c ||} 
 \hline
 Percentile Range & log-Likelihood & $\hat{\alpha}_{s}$ & $\hat{\alpha}_{sc}$ & $\hat{\alpha}_{cs}$ & $\hat{\alpha}_{c}$ \\ [0.5ex] 
 \hline\hline
 0.0-0.1 & 318.88 & 6.331 & 0.000 & 0.000 & 4.920 \\ 
 \hline
 0.1-0.2 & 288.02 & 6.180 & 0.018 & 0.040 & 1.653 \\
 \hline
  0.2-0.3 & 280.83 & 6.139 & 0.000 & 0.170 & 1.502 \\
 \hline
  0.3-0.4 & 262.59 & 5.167 & 0.030 & 0.094 & 2.808 \\
 \hline
  0.4-0.5 & 275.00 & 5.062 & 0.000 & 0.000 & 7.649 \\
  \hline
  0.5-0.6 & 339.13 & 3.467 & 0.000 & 0.000 & 5.535 \\
  \hline
  0.6-0.7 & 292.60 & 3.321 & 0.062 & 0.001 & 8.841 \\
  \hline
  0.7-0.8 & 296.92 & 3.293 & 0.000 & 0.000 & 7.160 \\
  \hline
  0.8-0.9 & 307.83 & 3.293 & 0.000 & 0.000 & 5.782 \\ 
 \hline
  0.9-1.0 & 382.86 & 3.688 & 0.000 & 0.000 & 9.611 \\ [1ex] 
  \hline
\end{tabular}
\begin{tabular}{|| c | c c c c ||} 
 \hline
 Percentile Range & $\hat{\beta}_{s}$ & $\hat{\beta}_{sc}$ & $\hat{\beta}_{cs}$ & $\hat{\beta}_{c}$ \\ [0.5ex] 
 \hline\hline
 0.0-0.1 & 9.141 & 7.242 & 9.192 & 6.119 \\ 
 \hline
 0.1-0.2 & 8.671 & 6.256 & 0.000 & 5.188 \\
 \hline
  0.2-0.3 & 8.554 & 5.504 & 1.259 & 5.382 \\
 \hline
  0.3-0.4 & 8.572 & 1.168 & 0.286 & 3.393 \\
 \hline
  0.4-0.5 & 8.409 & 10.0+ & 8.863 & 9.993  \\
  \hline
  0.5-0.6 & 6.649 & 6.574 & 2.509 & 5.928  \\
  \hline
  0.6-0.7 & 6.577 & 8.137 & 9.929 & 9.457 \\
  \hline
  0.7-0.8 & 6.536 & 9.999 & 9.616 & 7.611 \\
  \hline
  0.8-0.9 & 6.532 & 8.131 & 1.660 & 5.942 \\ 
  \hline
  0.9-1.0 & 6.705 & 8.081 & 2.466 & 10.0+ \\ [1ex] 
  \hline
\end{tabular}
\begin{tabular}{|| c | c c ||} 
 \hline
 Percentile Range & $\hat{\mu}_{s}$ & $\hat{\mu}_{c}$ \\ [0.5ex] 
 \hline\hline
 0.0-0.1 & 1.392 & 3.566 \\ 
 \hline
 0.1-0.2 & 1.320 & 8.561 \\
 \hline
  0.2-0.3 & 1.271 & 9.326 \\
 \hline
  0.3-0.4 & 1.285 & 0.771 \\
 \hline
  0.4-0.5 & 1.531 & 2.324 \\
  \hline
  0.5-0.6 & 1.585 & 0.419 \\
  \hline
  0.6-0.7 & 1.607 & 0.398 \\
  \hline
  0.7-0.8 & 1.650 & 0.362 \\
  \hline
  0.8-0.9 & 1.648 & 0.164 \\ 
  \hline
   0.9-1.0 & 1.510 & 0.267 \\ [1ex] 
  \hline
\end{tabular}
\end{center}
\section{\\Maximum Likelihood Estimation Code (Mathematica)}
\begin{doublespace}
\noindent\text{Matrix of arrival times for each time series (BTC and USDT).}\\
\noindent\(\pmb{t=}
\pmb{\{\{...\}-t_0,}
\pmb{\{...\}-t_0\};}\)
\end{doublespace}
\noindent\text{The below code is broken out explicitly for the 2-variate case.}
\newline
\noindent\text{Note that runtime of this construction is by no means maximally-efficient.}\\
\begin{doublespace}
\noindent\text{Begin by specifying the recursive functions in the log-Likelihood equation.}\\
\noindent\(\pmb{M = \text{Length}[t];}\\
\pmb{T=\text{Max}[\text{Table}[t[[i]][[-1]],\{i,M\}]];}\\
\pmb{}\\
\pmb{\text{R11}[\text{k$\_$},\text{b11$\_$}]\text{:=}\text{Piecewise}[}\\
\pmb{\{}\\
\pmb{\{0,k==0\}}\\
\pmb{\},}\\
\pmb{E{}^{\wedge}(-\text{b11}*(t[[1]][[k]]-t[[1]][[k-1]]))*\text{R11}[k-1,\text{b11}]+}\\
\pmb{\text{Piecewise}[}\\
\pmb{\{}\\
\pmb{\{0,\text{Length}[\text{Position}[t[[1]],\_?(t[[1]][[k-1]]\leq \#<t[[1]][[k]]\&)]]==0\}}\\
\pmb{\},}\\
\pmb{\text{Sum}[}\\
\pmb{E{}^{\wedge}(-\text{b11}*(t[[1]][[k]]-t[[1]][[i]])),}\\
\pmb{\{i,\text{Flatten}[\text{Position}[t[[1]],\_?(t[[1]][[k-1]]\leq \#<t[[1]][[k]]\&)]][[1]],}\\
\pmb{\text{Flatten}[\text{Position}[t[[1]],\_?(t[[1]][[k-1]]\leq \#<t[[1]][[k]]\&)]][[-1]]\}}\\
\pmb{]}\\
\pmb{]}\\
\pmb{]}\\
\pmb{}\\
\pmb{\text{R12}[\text{k$\_$},\text{b12$\_$}]\text{:=}\text{Piecewise}[}\\
\pmb{\{}\\
\pmb{\{0,k==0\}}\\
\pmb{\},}\\
\pmb{E{}^{\wedge}(-\text{b12}*(t[[1]][[k]]-t[[1]][[k-1]]))*\text{R12}[k-1,\text{b12}]+}\\
\pmb{\text{Piecewise}[}\\
\pmb{\{}\\
\pmb{\{0,\text{Length}[\text{Position}[t[[2]],\_?(t[[1]][[k-1]]\leq \#<t[[1]][[k]]\&)]]==0\}}\\
\pmb{\},}\\
\pmb{\text{Sum}[}\\
\pmb{E{}^{\wedge}(-\text{b12}*(t[[1]][[k]]-t[[2]][[i]])),}\\
\pmb{\{i,\text{Flatten}[\text{Position}[t[[2]],\_?(t[[1]][[k-1]]\leq \#<t[[1]][[k]]\&)]][[1]],}\\
\pmb{\text{Flatten}[\text{Position}[t[[2]],\_?(t[[1]][[k-1]]\leq \#<t[[1]][[k]]\&)]][[-1]]\}}\\
\pmb{]}\\
\pmb{]}\\
\pmb{]}\\
\pmb{}\\
\pmb{\text{R21}[\text{k$\_$},\text{b21$\_$}]\text{:=}\text{Piecewise}[}\\
\pmb{\{}\\
\pmb{\{0,k==0\}}\\
\pmb{\},}\\
\pmb{E{}^{\wedge}(-\text{b21}*(t[[2]][[k]]-t[[2]][[k-1]]))*\text{R21}[k-1,\text{b21}]+}\\
\pmb{\text{Piecewise}[}\\
\pmb{\{}\\
\pmb{\{0,\text{Length}[\text{Position}[t[[1]],\_?(t[[2]][[k-1]]\leq \#<t[[2]][[k]]\&)]]==0\}}\\
\pmb{\},}\\
\pmb{\text{Sum}[}\\
\pmb{E{}^{\wedge}(-\text{b21}*(t[[2]][[k]]-t[[1]][[i]])),}\\
\pmb{\{i,\text{Flatten}[\text{Position}[t[[1]],\_?(t[[2]][[k-1]]\leq \#<t[[2]][[k]]\&)]][[1]],}\\
\pmb{\text{Flatten}[\text{Position}[t[[1]],\_?(t[[2]][[k-1]]\leq \#<t[[2]][[k]]\&)]][[-1]]\}}\\
\pmb{]}\\
\pmb{]}\\
\pmb{]}\\
\pmb{}\\
\pmb{\text{R22}[\text{k$\_$},\text{b22$\_$}]\text{:=}\text{Piecewise}[}\\
\pmb{\{}\\
\pmb{\{0,k==0\}}\\
\pmb{\},}\\
\pmb{E{}^{\wedge}(-\text{b22}*(t[[2]][[k]]-t[[2]][[k-1]]))*\text{R22}[k-1,\text{b22}]+}\\
\pmb{\text{Piecewise}[}\\
\pmb{\{}\\
\pmb{\{0,\text{Length}[\text{Position}[t[[2]],\_?(t[[2]][[k-1]]\leq \#<t[[2]][[k]]\&)]]==0\}}\\
\pmb{\},}\\
\pmb{\text{Sum}[}\\
\pmb{E{}^{\wedge}(-\text{b22}*(t[[2]][[k]]-t[[2]][[i]])),}\\
\pmb{\{i,\text{Flatten}[\text{Position}[t[[2]],\_?(t[[2]][[k-1]]\leq \#<t[[2]][[k]]\&)]][[1]],}\\
\pmb{\text{Flatten}[\text{Position}[t[[2]],\_?(t[[2]][[k-1]]\leq \#<t[[2]][[k]]\&)]][[-1]]\}}\\
\pmb{]}\\
\pmb{]}\\
\pmb{]}\\
\pmb{}\\
\pmb{}\\
\noindent\text{Use the above in explicitly defining the \textbf{negative} log-Likelihood function in the 2-variate case.}\\
\pmb{\text{nLL2}[\text{a11$\_$}, \text{a12$\_$},\text{a21$\_$},\text{a22$\_$}, \text{b11$\_$}, \text{b12$\_$},\text{b21$\_$},\text{b22$\_$}, \text{m1$\_$},\text{m2$\_$}]\text{:=}
-(}\\
\pmb{T*(-\text{m1} -\text{m2})}\\
\pmb{-(\text{a11}/\text{b11}*\text{Sum}[1-E{}^{\wedge}(-\text{b11}*(T-t[[1]][[k]])),\{k,1,\text{Length}[t[[1]]]\}] + }\\
\pmb{\text{a12}/\text{b12}*\text{Sum}[1-E{}^{\wedge}(-\text{b12}*(T-t[[2]][[k]])),\{k,1,\text{Length}[t[[2]]]\}]+}\\
\pmb{\text{a21}/\text{b21}*\text{Sum}[1-E{}^{\wedge}(-\text{b21}*(T-t[[1]][[k]])),\{k,1,\text{Length}[t[[1]]]\}]+}\\
\pmb{\text{a22}/\text{b22}*\text{Sum}[1-E{}^{\wedge}(-\text{b22}*(T-t[[2]][[k]])),\{k,1,\text{Length}[t[[2]]]\}])}\\
\pmb{+\text{Sum}[\text{Log}[\text{m1}+\text{a11}*\text{R11}[k,\text{b11}] + \text{a12}*\text{R12}[k,\text{b12}]],\{k,1,\text{Length}[t[[1]]]\}]+}\\
\pmb{\text{Sum}[\text{Log}[\text{m2}+\text{a21}*\text{R21}[k,\text{b21}] + \text{a22}*\text{R22}[k,\text{b22}]],\{k,1,\text{Length}[t[[2]]]\}]}\\
\pmb{)}\)
\end{doublespace}
\noindent\text{Execute minimization procedure with domain constraints.} 
\newline
\noindent\text{Specify Nelder-Mead optimization algorithm.}\\
\begin{doublespace}
\noindent\(\pmb{\max  = 10;}\\
\pmb{\text{NMinimize}[}\\
\pmb{\{\text{nLL2}[\text{a11}, \text{a12},\text{a21},\text{a22}, \text{b11}, \text{b12},\text{b21},\text{b22}, \text{m1},\text{m2}], 0.000000000001\leq
\text{a11} <\max , }\\
\pmb{0.000000000001\leq \text{a12} <\max , 0.000000000001\leq \text{a21} <\max , 0.000000000001\leq \text{a22} <\max ,}\\
\pmb{0.000000000001\leq \text{b11}<\max , 0.000000000001\leq \text{b12} <\max , 0.000000000001\leq \text{b21} <\max , }\\
\pmb{0.000000000001\leq \text{b22} <\max ,0.000000000001\leq \text{m1} <\max , 0.000000000001\leq \text{m2}<\max \},}\\
\pmb{\{\text{a11}, \text{a12},\text{a21},\text{a22}, \text{b11}, \text{b12},\text{b21},\text{b22}, \text{m1},\text{m2}\},\text{Method}\to \text{{``}NelderMead{''}},
}\\
\pmb{\text{MaxIterations}\to 10000]}\)
\end{doublespace}

\end{document}